\documentclass[12pt]{article}
\usepackage{epsfig}
\usepackage{psfig}
\begin {document}

\hfill Astroparticle Phys., submitted 11 October 2002

\hfill revised 19 November 2002

\begin{center}
{\large TeV gamma rays and cosmic rays from the nucleus of M87,\\ a mis-aligned BL Lac object}\\[1cm]
{R.J.\ Protheroe$^{1,*}$, A.-C.\ Donea$^1$, A.\ Reimer$^2$\\
$^1$Department of Physics and Mathematical Physics, The University of Adelaide, Adelaide, SA 5005, Australia\\
$^2$Ruhr-Universit\"at Bochum, Institut f\"ur Theoretische Physik, Lehrstuhl IV: Weltraum- und Astrophysik, D-44780 Bochum, Germany}\\
\end{center} 

\begin{abstract}
The unresolved nuclear region of M87 emits strong non-thermal
emission from radio to X-rays.  Assuming this emission to
originate in the pc scale jet aligned at $\theta \sim 30^\circ$
to the line of sight, we interpret this emission in the context
of the Synchrotron Proton Blazar (SPB) model.  We find the
observed nuclear jet emission to be consistent with M87 being a
mis-aligned BL Lac Object and predict gamma-ray emission
extending up to at least 100~GeV at a level easily detectable by
GLAST and MAGIC, and possibly by VERITAS depending on whether it
is high-frequency or low-frequency peaked.  Predicted neutrino
emission is below the sensitivity of existing and planned
neutrino telescopes.  Ultra-high energy neutrons produced in pion
photoproduction interactions decay into protons after escaping
from the host galaxy.  Because energetic protons are deflected by the
intergalactic magnetic field, the protons from the decay of neutrons
emitted in all directions, including along the jet axis where the
Doppler factor and hence emitted neutron energies are higher, can
contribute to the observed ultra-high energy cosmic rays.  We
consider the propagation of these cosmic ray protons to Earth and
conclude that M87 could account for the observed flux if the
extragalactic magnetic field topology were favourable.
\end{abstract}

\noindent {PACS:}\\
98.54.Cm  Active and peculiar galaxies (including BL Lacertae objects,
          blazars, Seyfert galaxies, Markarian galaxies, and active
          galactic nuclei)\\
98.54.Gr  Radio galaxies\\
98.58.Fd  Jets, outflows and bipolar flows\\
98.70.Rz  gamma-ray sources\\
98.70.Sa  Cosmic rays (including sources, origin, acceleration, and
          interactions)\\
13.85.Tp  Cosmic-ray interactions\\

\noindent $^{*}${email: rprother@physics.adelaide.edu.au}

\section{Introduction}

M87 is usually classified as a Fanaroff-Riley Class I (FR-I)
radio galaxy having a one-sided relativistic jet \cite{Biretta99}
implying, in the context of unification models \cite{UrryPadovani95} that
its unresolved nuclear region is a misaligned BL Lac object.  The
viewing angle is estimated to be $\theta \approx 30^\circ$ giving
a maximum Doppler factor
$\delta=[\gamma_j(1-\beta_j\cos\theta)]^{-1}$ of $\delta_{\rm
max} \approx 2$ where $\gamma_j=(1-\beta_j^2)^{-1/2}$ is the jet
Lorentz factor.  For simplicity, we shall take $\delta=1$ but
consider the range $0.66 \le \delta \le 1.6$ for the M87 jet at
pc scales.  Although the jet emission is not strongly Doppler
boosted towards us -- it may even be de-boosted -- its proximity
to us (16.3 Mpc \cite{Cohen2000}) partially compensates for this,
and makes TeV gamma-ray \cite{BaiLee2001}, cosmic ray
\cite{Ahn2000,Biermanetal2001} and possibly neutrino emission from this object
potentially interesting.

The strong variability of the optical flux of M87 suggests that
the jet emits synchrotron radiation at optical frequencies
somewhere within an unresolved central region less than 5~pc in
diameter \cite{Tsvetanov1998}.  The jet  has well-defined
relativistic features, and it is remarkable that despite its low
power the jet extends beyond 30 arcsec from the core
\cite{Sparks1996}, possibly being a remnant from when M87 was
much more active.  Given the large black hole mass $3\times 10^9
$M$_{\odot}$, and the low jet luminosity $5 \times 10^{44}$ erg
s$^{-1}$ \cite{Owen2000}, the accretion disk must be currently
in a low radiative state and provide little power to the jet.  As
a consequence, the heating of any torus is currently inefficient
and would produce little attenuation of TeV gamma-ray signals
although this would change if M87 returned to an active state
commensurate with its high black hole mass \cite{DoneaProtheroe2002}.

BL Lac objects, along with flat-spectrum radio quasars are
collectively referred to as blazars.  BL Lacs may be
high-frequency or low-frequency peaked (HBLs and LBLs).  Their
broad-band spectra consist of two spectral components which
appear as broad `humps' in the spectral energy distribution (SED), and
are due to emission from a jet oriented at small angle with
respect to the line-of-sight.  The low-energy component, is
generally believed to be synchrotron emission from relativistic
electrons, and extends from the radio to UV or X-ray frequencies.
The origin of the high-energy component, starting at X-ray or
$\gamma$-ray energies and extending in some cases to TeV-energies, is
uncertain.  

For BL Lacs which have rather weak thermal emission, the favoured
``leptonic model'' is the so-called synchrotron-self Compton
(SSC) model in which the same relativistic electrons responsible
for the low energy synchrotron hump in the SED up-scatter
synchrotron photons to high energies via the Inverse Compton
effect.  ``Hadronic models'' were proposed more than 10 years ago
to explain the $\gamma$-ray emission from
blazars~\cite{MB89}. Recently M\"ucke \& Protheroe
\cite{MP2000,MP2001a} have discussed in detail the various
contributing emission processes in the synchrotron proton blazar
(SPB) model.  In hadronic models the relativistic jet consists of
relativistic proton and electron components, which again move
relativistically along the jet. High-energy radiation is produced
through photomeson production, and through proton and muon
synchrotron radiation, and subsequent synchrotron-pair cascading
in the highly magnetized environment.  In the case of BL Lacs
internal photon fields (i.e. produced by synchrotron radiation
from the co-accelerated electrons) serve as the target for pion
photoproduction.  These models can, in principle, be
distinguished from leptonic models by the observation of high
energy neutrinos generated in decay chains of mesons created in
the photoproduction interactions (for a recent review see
\cite{LM2000}).  

In hadronic models, AGN would contribute also to the pool of
extragalactic ultra-high energy cosmic rays (UHECR) through the
decay outside the host galaxy of neutrons produced by
photoproduction interactions (see \cite{ProtheroeSzabo1992} in
the context of accretion shocks, and \cite{Mannheimetal2002} in
the context of jets) or by escape of protons directly accelerated
at termination shocks of jets in giant radio lobes of
Fanaroff-Riley Class II (FR-II) radio galaxies
\cite{RachenBiermann1993}.  

Greisen \cite{Gre66}, and Zatsepin and Kuz'min \cite{Zat66} (GZK)
showed that the nucleonic component of UHECR above $10^{20}$ eV
will be severely attenuated in the cosmic microwave background
radiation (CMBR), primarily due to pion photoproduction
interactions.  At $3\times 10^{20}$~eV the mean free path is
$\sim 5$~Mpc and the energy-loss distance is $\sim
20$~Mpc~\cite{Stecker68} (see Stanev et al.~\cite{Stanevetal2000}
for recent calculations).  Thus, if the UHECR are extragalactic,
one would expect their spectrum to cut off at $\sim 10^{20}$ eV,
the ``GZK cut-off''.  UHECR may have been observed with energies
well above $10^{20}$ eV (see Nagano and
Watson~\cite{NaganoWatson2000} for a review including a
discussion of models for the origin of the UHECR).  However, very
recent data from the two largest aperture high energy cosmic ray
detectors are contraditory: AGASA~\cite{AGASA2002} observes no
GZK cut-off while HiRes~\cite{HiRes02} observes a cut-off
consistent with the expected GZK cut-off.  A systematic
over-estimation of energy of about 25\% by AGASA or
under-estimation of energy of about 25\% by HiRes could account
the discrepancy ~\cite{HiRes02}, and the continuation of the
UHECR spectrum to energies well above $10^{20}$ eV is now far from
certain.  Nevertheless, if the spectrum does extend well beyond
$10^{20}$ eV, even though it is only an FR-I radio galaxy M87 is
an attractive possibility for the origin of the UHECR
\cite{Ahn2000,Biermanetal2001} because of its proximity.

\section{SPB modelling M87 as a mis-aligned BL Lac}

\begin{figure} 
\centerline{\psfig{file=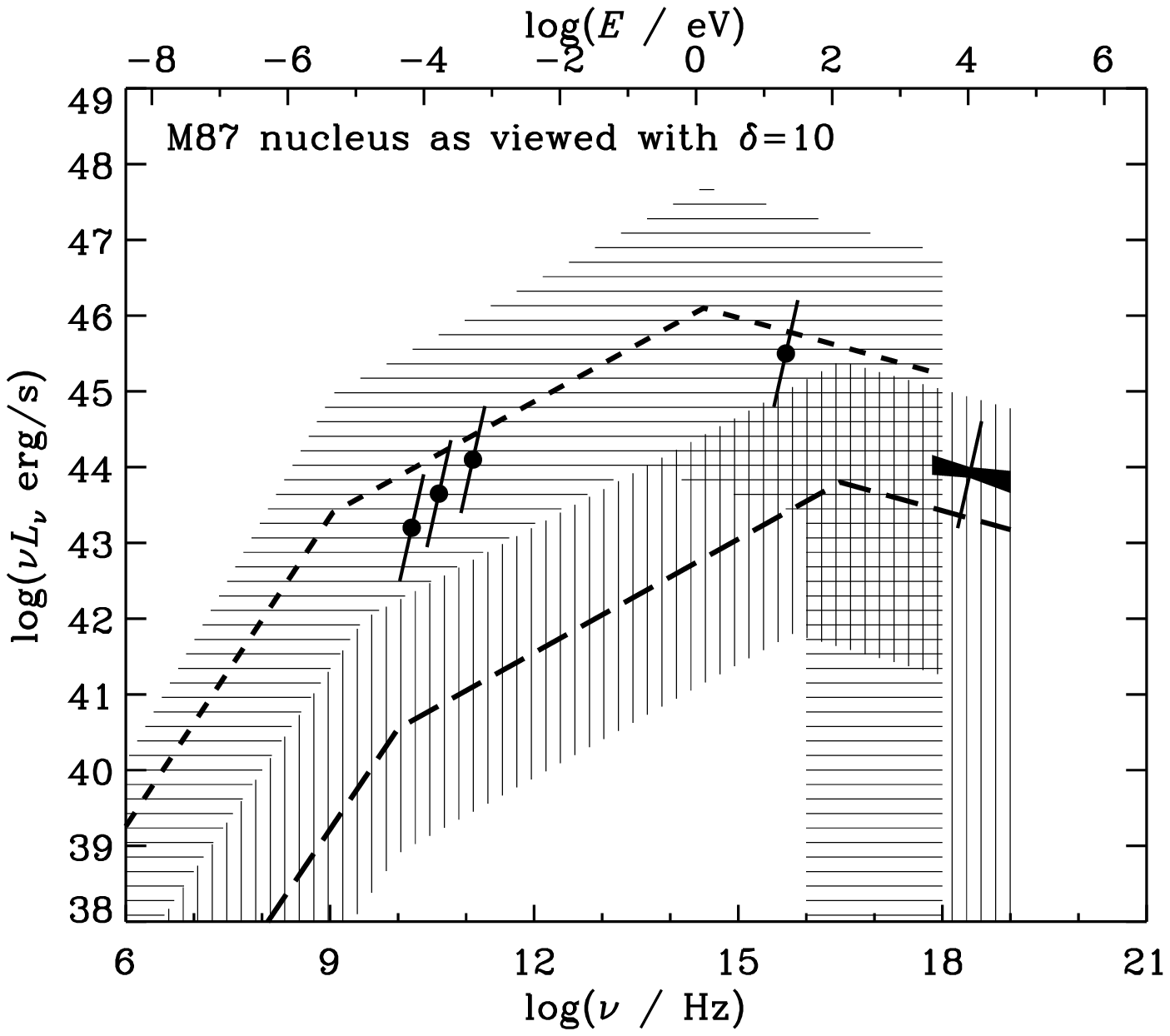,width=\hsize}} 
\caption{Form of the SED assumed for the synchrotron radiation
from LBLs (short dashed curves) and HBLs (long dashed curves).
The horizontal shading encompasses the SEDs of all LBLs, and the
vertical shading encompasses the SEDs of all HBLs considered by
Ghisellini et al.~\protect\cite{Gh98}.  Data from M87 unresolved
nuclear jet emission
refs.~\protect\cite{BirettaSternHarris91,Perlman2001,WilsonYang2002}
Doppler boosted to $\delta=10$ have been added (see text for
details).  Error bars correspond to uncertainty in Doppler
boosting due to uncertainty in Doppler factor of M87 which we
taken to be in the range $0.67 \le \delta \le 1.5$.}
\label{Fig1}
\end{figure}

M\"ucke et al.~\cite{Muecke_etal2002} identified the critical parameters
determining the properties of BL Lacs in the context of the
SPB model and constructed the ``average'' synchrotron spectrum
for each class, HBLs and LBLs, which served as the target photon
distribution for their hadronic cascade.  An extensive collection
of blazar SEDs published by Ghisellini et al. \cite{Gh98} were
used to construct the ``average'' SED of HBLs and LBLs by
overlaying all available HBL and LBL SEDs.  They found that a
broken power law gave a reasonable representation of the
low-frequency hump in the SED of both HBLs and LBLs
\begin{equation}
n(\epsilon) \propto \left\{ \begin{array}{ll} \epsilon^{-\alpha_1} &
 \mbox{for} \qquad  \epsilon_i\leq \epsilon \leq
 \epsilon_b\nonumber\\
\epsilon^{-\alpha_2} & \mbox{for}
 \qquad \epsilon_b \leq \epsilon \leq \epsilon_c
\end{array} \right.
\end{equation}
with $\alpha_1=1.5$, $\alpha_2=2.25$ and $\epsilon_i=10^{-5\ldots
-6}$eV.  The break energy $\epsilon_b$ of LBLs varied between
$\approx$ 0.1 eV to 1.3 eV, while the maximum synchrotron photon
energy $\epsilon_c$ ranged over two orders of magnitude, from
$\approx$ 41 eV to 4 keV. The populated energy range of HBLs is
more restricted: $\epsilon_b\approx$ 26 eV to 131 eV and
$\epsilon_c\approx$ 4.1 keV to 41 keV. The peak of the low-energy
SED was $\log{\nu L_{\nu}^{\rm{max}} {\rm (erg/s)}} \approx
45.6-46.1$ for LBLs and $\log{\nu L_{\nu}^{\rm{max}} {\rm
(erg/s)}} \approx 43.4-43.8$ for HBLs. M\"ucke et
al.~\cite{Muecke_etal2002} defined the ``average LBL'' by
$\epsilon_b=$1.3 eV, $\epsilon_c=$4.1 keV and $\log\nu
L_{\nu}^{\rm max} \mbox{  (erg/s)} = 46.1$, and the ``average
HBL'' by $\epsilon_b=$131 eV, $\epsilon_c=$41 keV and $\log{\nu
L_{\nu}^{\rm{max}}} = 43.8$.  The parametrizations for HBLs and
LBLs are visualized in Fig.~\ref{Fig1}, with the dashed lines
showing the ``average'' SED and the hatched regions showing the
range of the SEDs for the two cases.

Assuming that the unresolved nuclear emission in M87, i.e.\ from
the pc scale jet, has $\delta=1$ we have added to Fig.~\ref{Fig1}
the radio \cite{BirettaSternHarris91}, infrared
\cite{Perlman2001} and X-ray \cite{WilsonYang2002} data on the
nuclear region of M87 Doppler boosted by $\delta=10$ to mimic how
M87 would appear if its jet were closely aligned towards us such
that it would be classed as a BL Lac rather than a FR-I.  Since
the Doppler factor of M87 is not well known, taking the range
from 0.66 to 1.5, the observed SED as viewed along the axis with
$\delta=10$ would move along the error bars added to the data.
From Fig.~\ref{Fig1}, the rather high radio-infrared luminosity
would suggest that M87 could be a mis-aligned LBL, in which case
the X-ray emission would be part of the high-energy hump in the
SED due to cascading initiated by photons and electrons from pion
decay or by synchrotron photons emitted by protons and muons.
However, the X-ray emission is at a level closer to that expected
from an HBL, in which case it would be mainly due to synchrotron
emission by directly accelerated electrons.  One should bear in
mind that there is considerable variation in the SEDs of HBLs and
LBLs, and so one should not expect the SED of an individual
BL Lac to be identical to our ``average'' spectrum.  This is
particularly true for the M87 nuclear region where the
observations at radio, infrared and X-rays were not simultaneous
and are known to vary.  In addition, Wilson and Yang
\cite{WilsonYang2002} have calculated from the Chandra X-ray
flux and spectrum observed from the inner part of the jet that
there is intrinsic absorption by cold matter with an equivalent
hydrogen column density of $3$--$5 \times 10^{20}$~cm$^{-2}$ in
M87, and so the unattenuated X-ray flux could be higher than
plotted.  Thus we shall be satisfied if the observed SED is
within a factor $\sim 2$ of an ``average'' SED.

\begin{figure} 
\centerline{\psfig{file=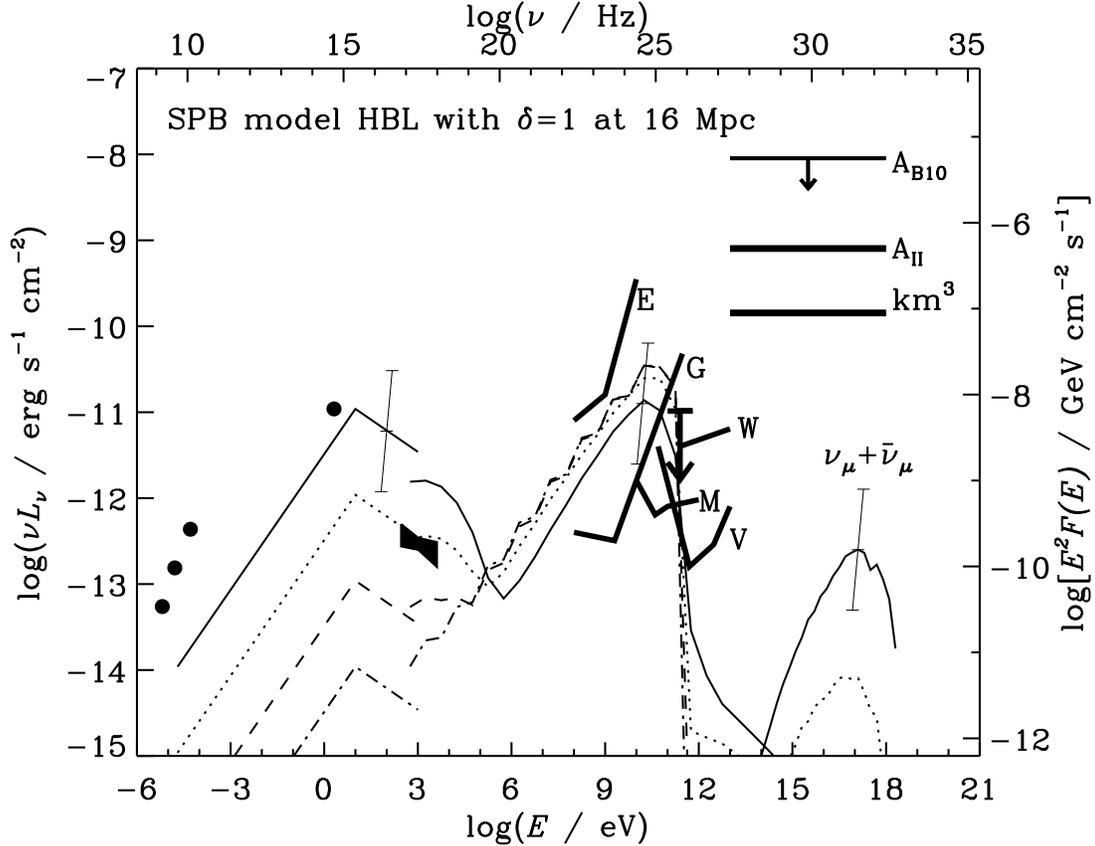,width=\hsize}} 
\caption{SED of emerging cascade radiation for different target
photon spectra (broken power-laws shown), $u'_B = u'_P$, $B = 30
G$, $\delta=1$, and $R' = 5\times 10^{15}$ cm.  HBL-like
synchrotron spectra with $u'_B = u'_P$, and
$\log(u'_{\rm{phot}}/\rm{eV cm}^{-3}) = 8$ (chain curves), 9
(dashed curves), 10 (dotted curves) and 11 (solid curves).  The
broken power-laws on the left show the electron synchrotron
radiation which provides the target photons for proton
interactions, the curves in the range $10^3$--$10^{14}$~eV show
the X-ray to gamma-ray flux due to proton interactions and proton
synchrotron radiation and subsequent cascading, and the curves in
the range $10^{14}$--$10^{18}$~eV show the corresponding neutrino
fluxes (note that neutrino fluxes for $\log(u'_{\rm{phot}}/\rm{eV
cm}^{-3}) = 8$ and 9 are too low to be included in this plot).
Error bars attached to solid curves correspond to uncertainty due
to uncertainty in Doppler factor of M87 which we taken to be in
the range $0.67 \le \delta \le 1.5$.  The sensitivities of the
EGRET (E), Whipple (W), GLAST (G), MAGIC (M) and VERITAS (V)
gamma ray telescopes are indicated, as are the sensitivities of
the AMANDA-II (A$_{\rm II}$) and IceCube (km$^3$) neutrino
telescopes, and an upper limit from AMANDA-B10 (A$_{\rm B10}$) --
see text for details.  Data from M87 unresolved nuclear jet
emission
refs.~\protect\cite{BirettaSternHarris91,Perlman2001,WilsonYang2002}
and the upper limit at 250~GeV from Whipple
\protect\cite{WhippleM87} have been added.}
\label{Fig2}
\end{figure}

In Fig.~\ref{Fig2} we show the SEDs of HBLs modelled by M\"ucke
et al.\ \cite{Muecke_etal2002} in the context of the SPB model as
they would appear if observed at $\delta=1$ as FR-I radio
galaxies at a distance of 16~Mpc appropriate to M87.  Results
taken from M\"ucke et al.~\cite{Muecke_etal2002} for four peak
luminosities of the low energy hump of the SED are shown after
shifting to $\delta=1$.  As we see, there is little variation in
the high energy hump which peaks at $\sim 10$--$100$~GeV at a
level just below the sensitivities of EGRET and the Whipple High
Energy Gamma Ray Telescope, but well above the sensitivity of
GLAST and future northern hemisphere large-area atmospheric
Cherenkov telescopes (ACT) such as MAGIC and VERITAS
(sensitivities for all these telescopes taken from
ref.~\cite{GLAST} are shown together with the upper limit from
Whipple \cite{WhippleM87}).  One should bear in mind, however,
that the uncertainty in the Doppler factor of M87 will give rise
to a much larger uncertainty in the bolometric luminosity.  These
uncertainties are indicated in the theoretical SED and neutrino
spectra by the slanted error-bars.  We have extended the work of
M\"ucke et al.~\cite{Muecke_etal2002} by calculating the neutrino
output for these four cases and plot the expected flux of
muon-neutrinos ($\nu_\mu$+$\bar{\nu}_\mu$). The highest
luminosity modelled for the low-energy hump of the SED (solid
curve) is in closest agreement with the M87 SED and predicts the
highest neutrino flux.  We have added an upper-limit for the
AMANDA-B10 neutrino telescope by taking the diffuse neutrino
background limit \cite{Ahrens2002amanda} and multiplying by
$2\pi$~sr (neutrino telescopes detect upward-going neutrinos).
Sensitivities calculated by Albuquerque et
al.~\cite{Albuquerque2001} of AMANDA-II and IceCube to $E^{-2}$
diffuse intensity, again multiplied by $2\pi$~sr to convert to
point source sensitivity have also been added.  Unlike the
100~GeV gamma-ray emission which should easily be detected by
future telescopes, the neutrino flux is well below detection
levels of future neutrino telescopes.

\begin{figure} 
\centerline{\psfig{file=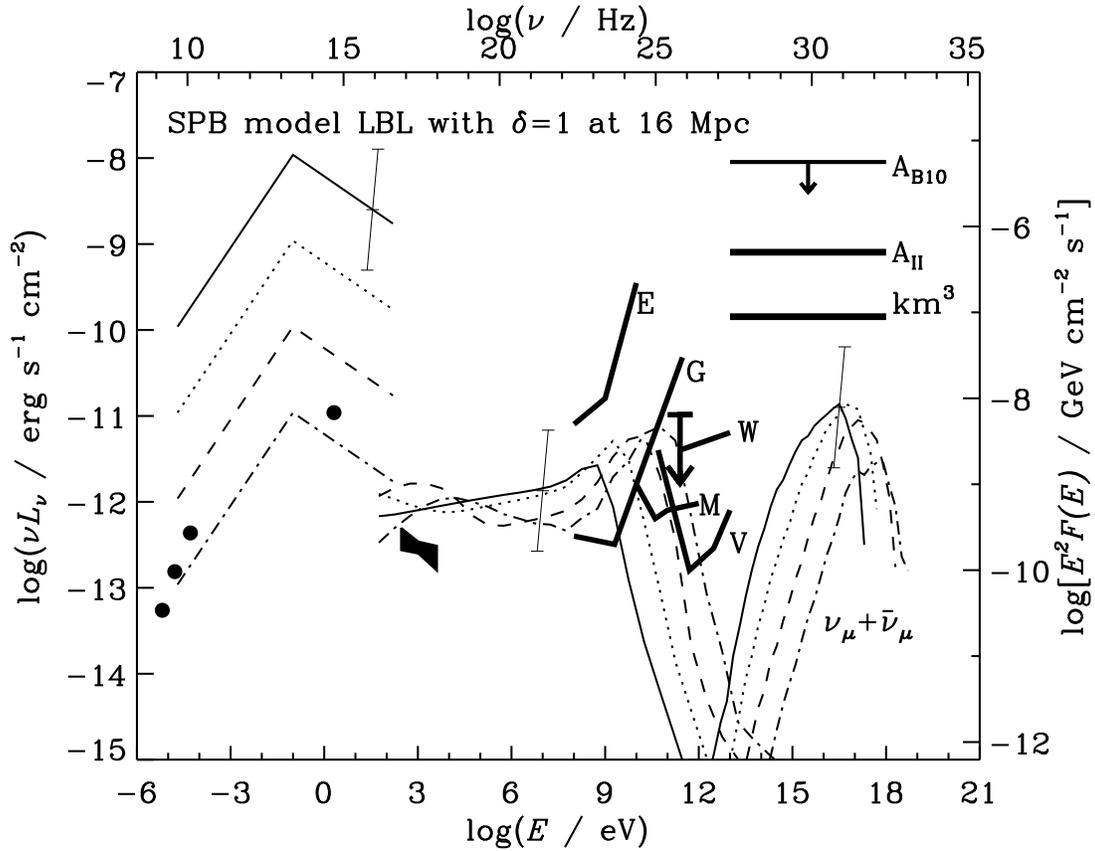,width=\hsize}} 
\caption{SED of emerging cascade radiation for different target
photon spectra (broken power-laws shown), $u'_B = u'_P$, $B = 30
G$, $\delta=1$, and $R' = 5\times 10^{15}$ cm.  LBL-like
synchrotron spectra with $\log(u'_{\rm{phot}}/\rm{eV cm}^{-3}) =
11$ (chain curves), 12 (dashed curves), 13 (dotted curves) and 14
(solid curves).  See Fig.~\protect\ref{Fig2} for key to other symbols.}
\label{Fig3}
\end{figure}

In Fig.~\ref{Fig3} we model M87 as a mis-aligned LBL in the
context of the SPB model.  In this case, we see that the
gamma-ray emission cuts off at 10~GeV--1~TeV depending on the
luminosity of the low energy hump of the SED.  This is because
for the higher target photon densities in LBL, pion
photoproduction losses determine the maximum proton energy, and
hence the maximum gamma-ray energy which results from the
cascades initiated by proton synchrotron radiation and pion decay
(including pion and muon synchrotron radiation).  This also
affects the maximum neutrino energy in the same way.  The lowest
luminosity modelled for the low-energy hump of the SED (chain
curve) is in closest agreement with the M87 SED and predicts the
highest gamma-ray and neutrino energies.  The gamma ray flux is a
factor $\sim 3$--10 lower than for the HBL case, but is still
detectable by future telescopes.  As with the HBL case, the
neutrino flux is below the sensitivity of IceCube.  However, it
is not impossible that in the case of rapid flaring, M87 might
just be detected as it is at a declination of $\sim +$$12^\circ$
and is ideally located for observation by a giant neutrino
telescope to be located at the South Pole such as IceCube because
it is sufficiently below the horizon to eliminate cosmic ray
events, while not being so far below the horizon that neutrino
absorption by the Earth's core becomes important.

\section{M87 as a source of UHECR}

Energetic protons magnetically trapped in AGN jets lose energy
predominantly by Bethe-Heitler pair production and pion
photoproduction on ambient radiation fields, or by adiabatic
deceleration as the jet expands.  Neutron production in pion
photoproduction sources (e.g. $p\gamma \to n\pi^+$) provides a
mechanism for escape of cosmic rays from the jet.  Neutrons decay
typically after travelling $(E_n/10^{20}$eV)~Mpc, which for UHECR
is well outside the host galaxy.  Recently Ahn et
al.~\cite{Ahn2000} and Biermann et al.~\cite{Biermanetal2001} have
shown that in the assumption of a Parker-spiral galactic wind magnetic structure out to $\sim 1.5$~Mpc, the
galactic wind of our Galaxy poses no restriction to the
entry to our Galaxy of UHECR from the general direction of M87,
and may even lead to clustering in the arrival directions of
cosmic rays from M87 observed from Earth.  These preferred arrival
directions and clustering may, however, be mainly due to the
Galactic wind topology and contain little information about the
direction of cosmic ray sources
\cite{BilloirLetessier-Selvon2000}.

\begin{figure} 
\centerline{\psfig{file=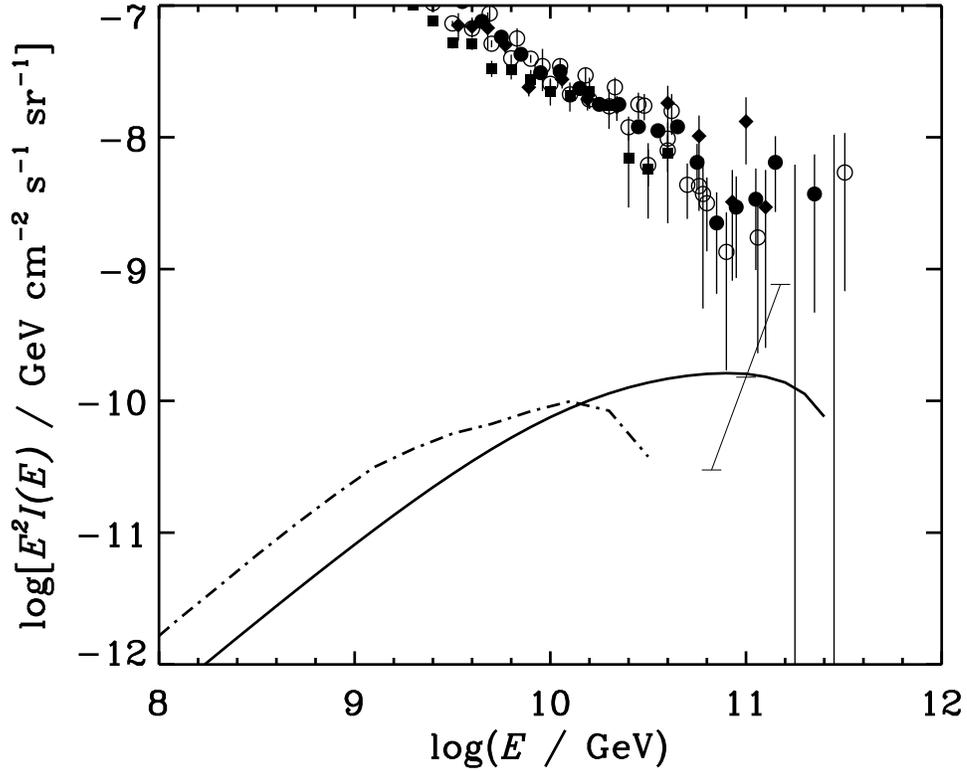,width=\hsize}} 
\caption{The observed intensity of UHECR taken from
ref.~\protect\cite{GaisserStanevCRreview}.  Chain curve gives
neutron flux divided by $4\pi$~sr (to convert flux to average
intensity) that would be observed at Earth from M87 if the
neutrons did not decay (intensity shown corresponds to LBL model
indicated by chain curve in Fig.~\protect\ref{Fig3}).  The solid
curve is the cosmic ray intensity approximation given by
Eq.~\protect\ref{eq:CRintensity} for this case (see text for a
discussion of diffusion and energy-loss in the CMBR not
included).  }
\label{Fig4}
\end{figure}

The observed intensity of UHECR \cite{GaisserStanevCRreview} (multipled by energy squared) is
plotted in Fig.~\ref{Fig4}.  For the most promising LBL model
(chain curve in Fig.~\ref{Fig3}) we plot (chain curve) the flux
of neutrons divided by $4\pi$~sr (to convert flux to average
intensity) that would be observed at Earth from M87 if the
neutrons did not decay (which of course they do).  Since these
neutrons would travel to Earth in straight lines, and since we
take M87 to have $\delta(\theta \sim 30^\circ)=1$, the maximum
neutron energy that would arrive at Earth would be approximately
the maximum jet-frame energy of the accelerated protons, which
would be $E_n = 3 \times 10^{19}$~eV for the model shown.
However, neutrons decay into protons whose directions are
isotropised in the intergalactic magnetic field before and during
propagation to Earth.  Hence we would have contributions to the
protons arriving at Earth from neutrons emitted at all angles
with respect to the jet axis, and hence from the full range of
Doppler factors corresponding to the jet Lorentz factor,
i.e.\ approximately $1/2\gamma_j \to 2\gamma_j$.  For example, if
the jet Lorentz factor were $\gamma_j=5$ a neutron with jet-frame
energy $E_n'=3 \times 10^{19}$~eV emitted along the jet axis
($\theta=0^\circ$) would decay to a proton with galaxy-frame energy $E
\approx 3 \times 10^{20}$~eV.  What we would observe therefore
depends on the galaxy-frame angle-averaged neutron luminosity
{\em on emission}, and would also have contributions from both jets
(i.e.\ twice that for one jet).  Given that the chain curve in
Fig.~\ref{Fig4} is for $\delta(\theta \sim 30^\circ) = 1$, it is
related directly to the jet-frame bolometric luminosity ${E_n'}^2
\dot{N}_n'(E_n')$.  For electromagnetic radiation and neutrinos
the {\em observed} bolometric luminosity is $\nu L_\nu^{\rm
obs}(\theta) =\delta^4(\theta) \nu'L'_{\nu'}$ and the angle-averaged
{\em emitted} bolometric luminosity is $\nu L_\nu^{\rm em} =
(4\pi)^{-1}\oint \delta^3(\theta) \gamma_j^{-1} \nu'L'_{\nu'}
d\Omega$.  Hence, by analogy, for ultra-relativistic protons of
energy $E \approx E_n$ from neutron decay, and remembering that
both jets contribute,
\begin{equation}
E^2 \dot{N}(E) = {2 \over 4\pi}\oint \delta^3(\theta)\gamma_j^{-1} 
{E_n'}^2 \dot{N}'_n(E_n')  d\Omega .
\end{equation}
We have added to Fig.~\ref{Fig4} (solid curve) the resulting cosmic ray
flux divided by $4\pi$~sr assuming $\gamma_j=5$ and straight-line 
propagation, i.e.
\begin{equation}
E^2 I(E) = {E^2 \dot{N}(E) \over (4\pi)4\pi d^2 }
\label{eq:CRintensity}
\end{equation}
corresponding to number density
\begin{equation}
n(E) = {\dot{N}(E) \over 4\pi d^2 c}.
\end{equation}
At $5 \times 10^{19}$--$3 \times 10^{20}$~eV this is a factor $\sim 20$ below
the observed UHECR.

Of course, so far this neglects effects of diffusive propagation
to Earth and interactions with the CMBR.  We shall consider first
the effects of diffusion.  For the case of a constant injection
rate of cosmic rays, and an infinite homogeneous diffusing medium
with diffusion coefficient $D(E)$ the number density at distance
$d$ is simply
\begin{equation}
n(E) = {\dot{N}(E) \over 4\pi D(E)d}.
\end{equation}
If we take $D(E)=\lambda (E) c/3$ then the observed intensity
will be enhanced with respect to straight-line propagation
by a factor 
\begin{equation}
g(E)=dc/D (E) = 3d/\lambda (E).
\end{equation}
For a  simple model of intergalactic space with a low magnetic field
($\sim 10^{-9}$~G) and a cell size $L_C > 100$~kpc, taking
$\lambda (E) \sim L_C$ would give an enhancement factor up to
$g(E) < 500$, such that the predicted UHECR from M87 could explain the
observed intensity above $10^{19}$~eV.  While diffusion appears to 
increase the cosmic ray intensity from M87, it also increases the 
travel time or effective distance travelled from M87 to Earth, 
\begin{equation}
d_{\rm eff} = d^2c/2D(E) = d g(E)/2.
\end{equation}
Hence for an enhancement factor of $g=20$ the effective distance
is $d_{\rm eff}=160$~Mpc.  At $10^{19}$~eV the mean energy-loss
distance is  approximately 1~Gpc and at $10^{20}$~eV the mean energy-loss
distance is approximately 160~Mpc, and so in this case there would only
be a significant energy loss above $10^{20}$~eV.

\begin{figure*} 
\centerline{\epsfig{file=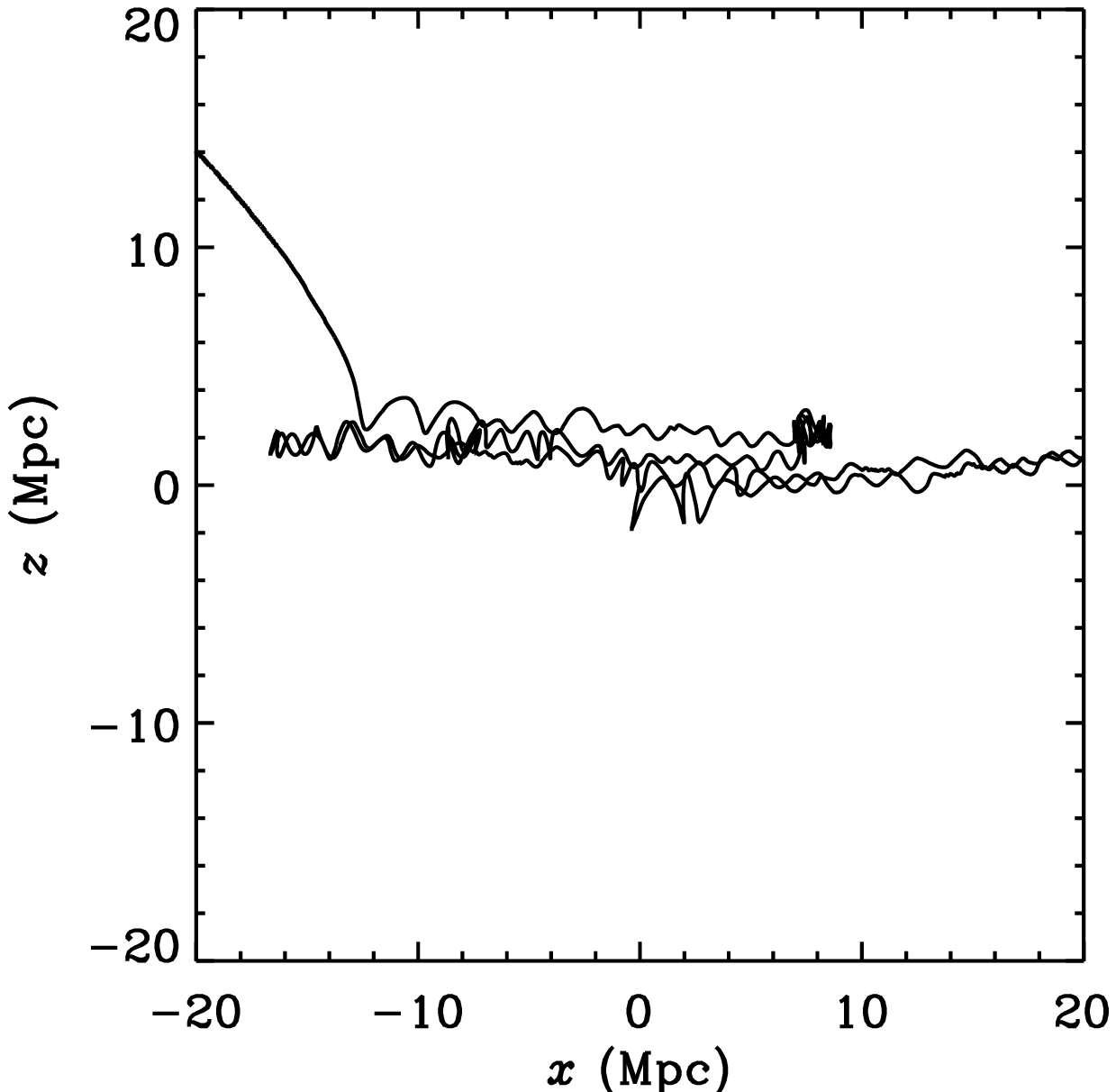,width=12cm}\hspace*{-2cm}\epsfig{file=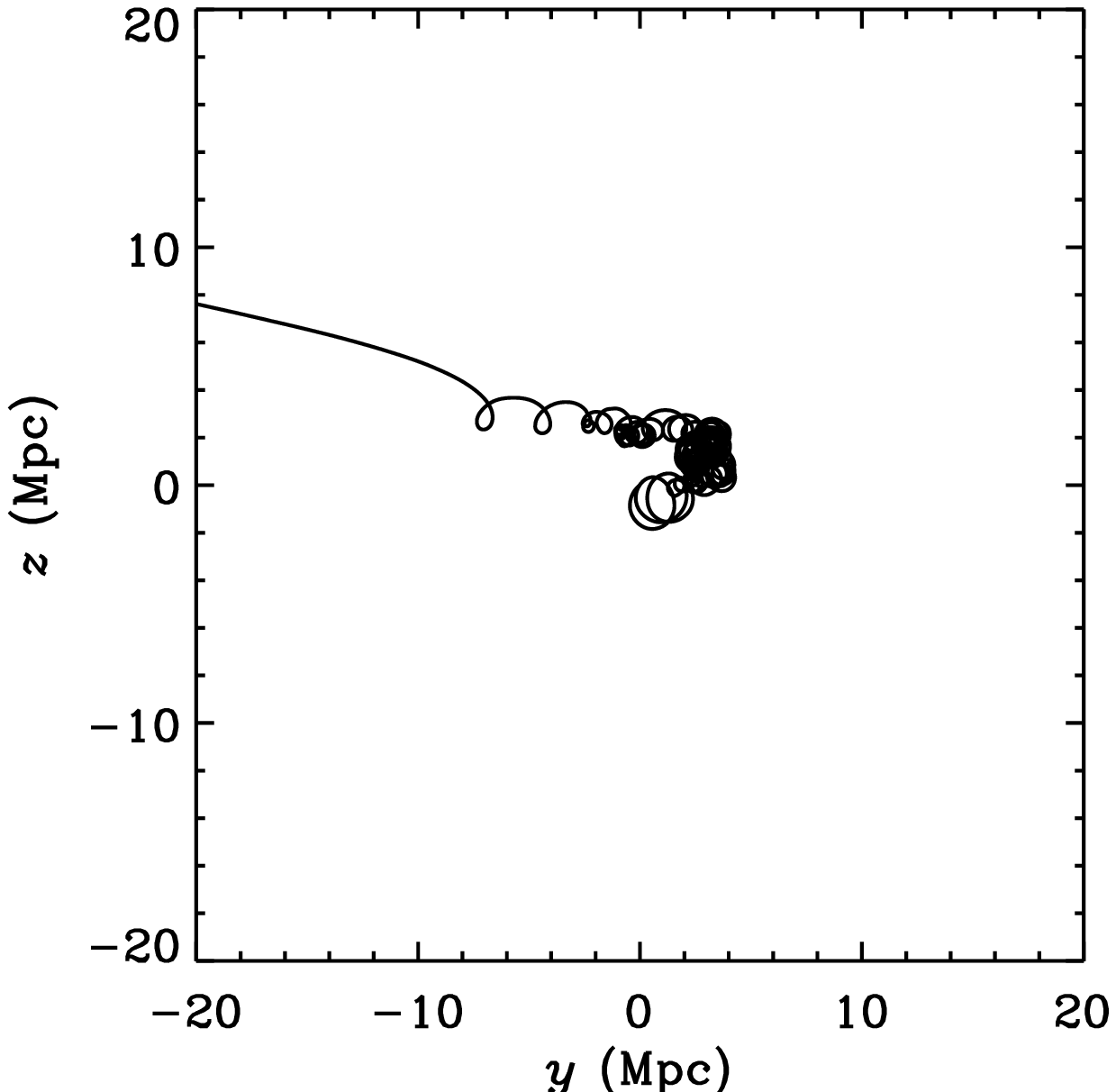,width=12cm}} 
\hspace*{-3.1cm}
\epsfig{file=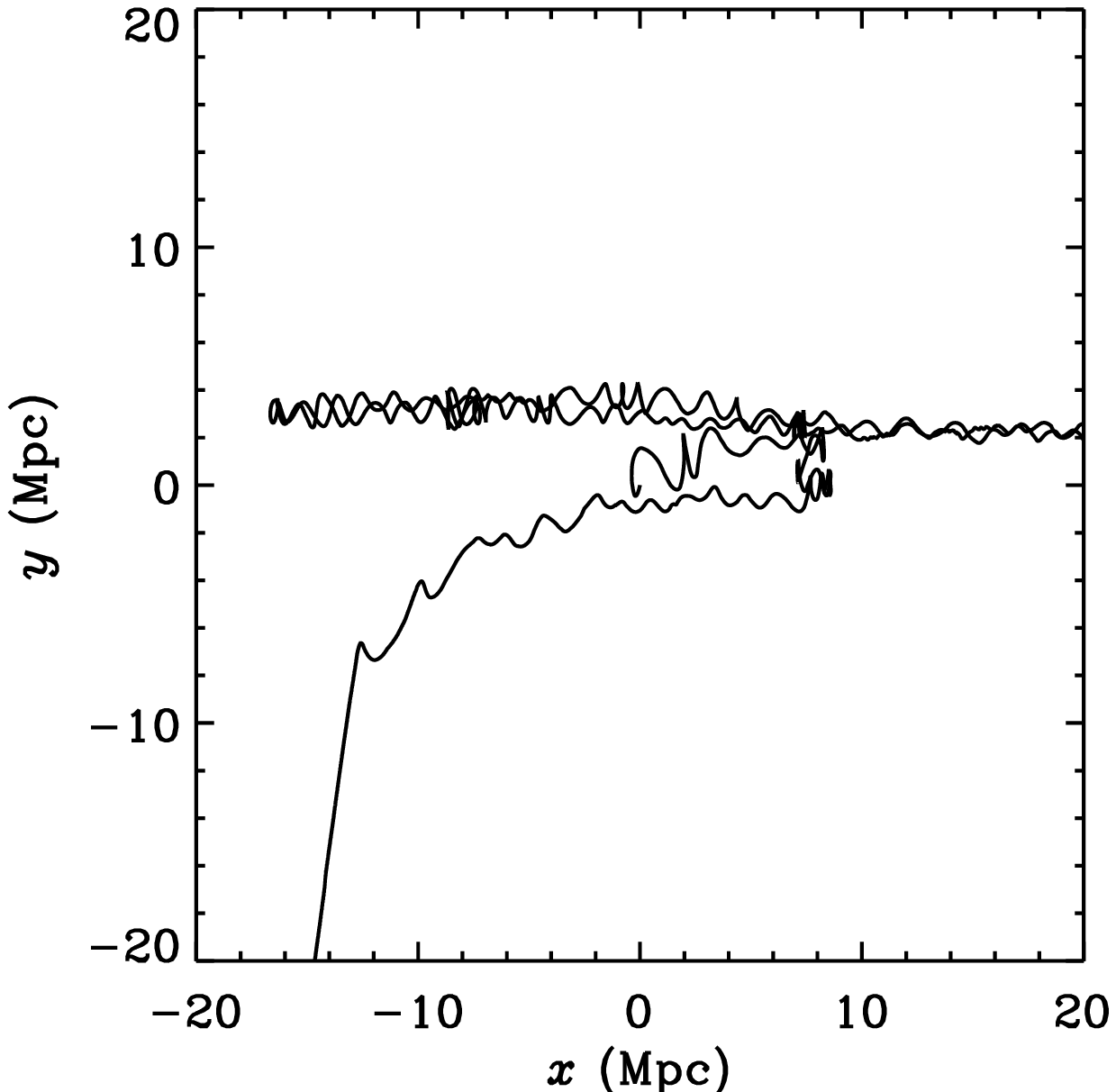,width=12cm}
\caption{Three orthogonal views of a typical trajectory of a
$10^{20}$~eV proton in the wall/void model of the IGMF discussed
in the text.  Particle is injected at the origin, regular
component of the magnetic field is in $x$ direction.  Wall
extends from $z=-2.5$~Mpc to $z=2.5$~Mpc. }
\label{Fig5}
\end{figure*}

The situation could actually be quite different if we consider an
intergalactic magnetic field (IGMF) structure based on the
observation of microgauss fields in clusters of galaxies
\cite{Kronberg1994}, and of clusters occurring in networks of
``walls'' separated by ``voids''.  It would then be reasonable to
expect relatively high fields of $10^{-7}$--$10^{-6}$~G in the
walls, and much lower fields, $10^{-11}$--$10^{-9}$~G, in the
voids.  Recently, Stanev et al.~\cite{StanevSeckelEngel2001} have
considered propagation of UHECR in three different models of the
IGMF in the local supergalactic structure in which the regular
field in the high-field region was $10^{-8}$~G.  We consider
propagation in a simple wall/void model similar to that used by Medina Tanco~\cite{MedinaTanco1998} to illustrate how UHECR would be deflected
in complex IGMF structures.  As M87 is at the centre of the Virgo
Cluster and is close to the super-galactic plane we assume it is
embedded in the higher field region.  To illustrate the effect of
propagation from M87 in a wall/void type IGMF we define the
origin of coordinates to be at M87 with the mid-plane of the wall
corresponding to the $x$--$y$ plane, the wall occupying $|z| <
2.5$~Mpc and the void occupying $|z| >7 .5$~Mpc.  Following
Medina Tanco~\cite{MedinaTanco1998} we adopt a regular magnetic field of
$10^{-7}$~G in the $x$-direction in the wall and $10^{-10}$~G in
the void, with a transition region sandwiched between the wall
and the void in which the magnetic field drops exponentially from
$10^{-7}$~G to $10^{-10}$~G.  The magnetic field contains an
irregular component having $\langle |\vec{B}_{\rm irreg}|^2
\rangle^{1/2}$ equal to 30\% of the regular component, and has a
Kolmogorov spectrum of turbulence.  The irregular field is
modelled as described in ref.~\cite{Stanevetal2000}
with wavenumbers corresponding to turbulence scales $L_C/2^n$
with $L_C=2.5$~Mpc and $n=0,\dots ,3$.  Particles are injected
isotropically at the origin and energy losses due to
Bethe-Heitler pair production and pion photoproduction are
included.  A typical simulation for $E_0=10^{20}$~eV is shown in
Fig.~\ref{Fig5}.

Whereas Medina Tanco \cite{MedinaTanco1998} was most interested
in the angular deflection of arriving cosmic rays relative to the
direction of the cosmic ray source, we are more interested in the
travel times of cosmic rays to Earth from the distance $d \approx
16$~Mpc to M87, and the total amount of time spent within unit
volume at distance $d$ from M87.  We divide the spherical shell
corresponding to 15~Mpc~$ <d< 17$~Mpc into 2160 equal cells of
volume $V$ with a grid in spherical coordinates $(\theta,\phi)$
using 72 equally spaced $\phi$-values and 30 equally spaced
$\cos\theta$-values.  Particle orbits are advanced in time steps
of size $\Delta t =0.03$~Mpc/$c$ chosen to be much smaller than
the smallest magnetic structure simulated.  At each step, the
proton energy is reduced according to the average energy-loss
rate (see e.g.\ figure~1 of ref.~\cite{Stanevetal2000}).  A check
is made at the end of each time step to determine whether the
proton is within the spherical shell, and if it is then $\Delta
t$ is added to the total time spent in the cell corresponding to
its spherical coordinates $(\theta,\phi)$ -- i.e.\ if on $k$
occasions a particle was found to be located inside the $i$th
cell at the end of a time step, then it would have spent
approximately a total time $k\Delta t$ inside the $i$th cell
since being injected at the origin.  Dividing by the number of
simulations one obtains the average time spent in a particular
cell per injected proton $\tau(\theta,\phi)$.  Similarly, the
time spent travelling since injection at M87 is added to the ages
of the protons ``detected'' in that cell.  Dividing by the total
number of occasions a particle was ``detected'' in that cell
gives the average age $\bar{t}(\theta,\phi)$ of the protons
present in that cell.

\begin{figure*} 
\centerline{\psfig{file=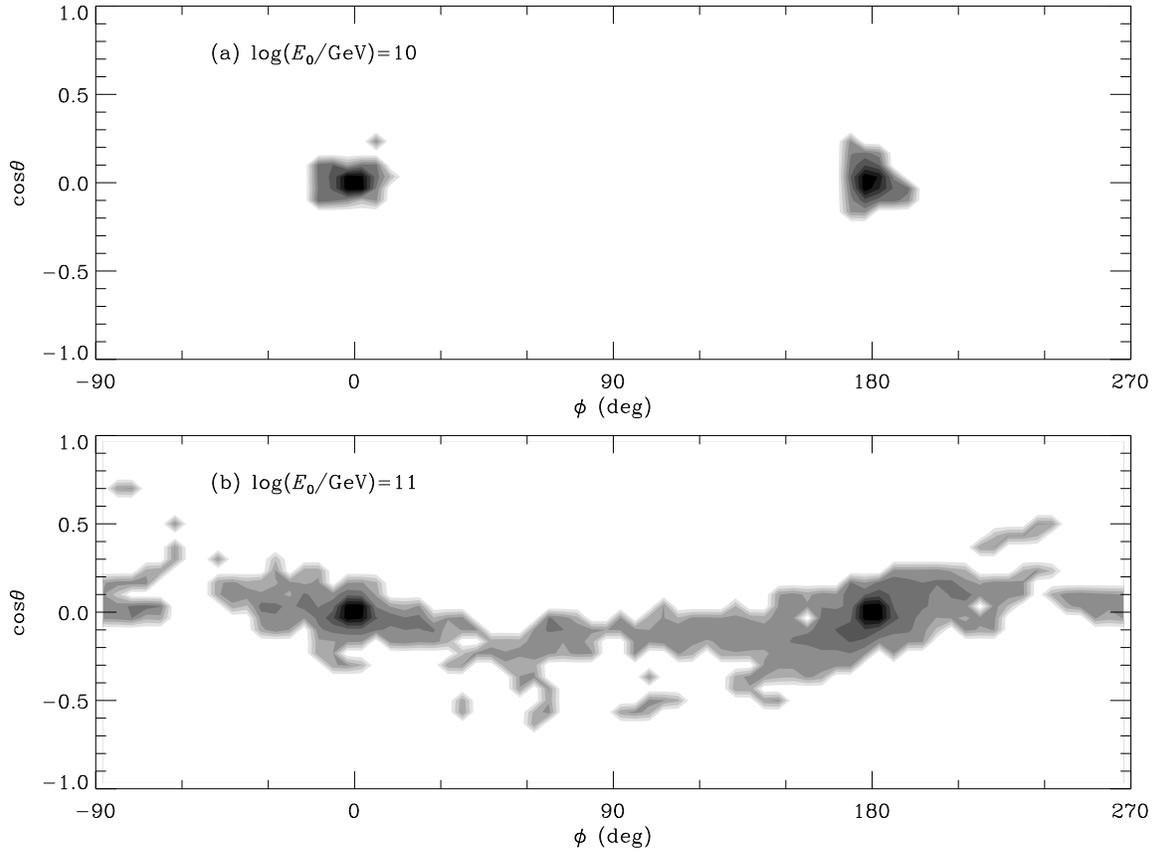,width=\hsize}} 
\caption{Enhancement factor $g(E)$ (Eq.~\protect\ref{eq:enhance})
as a function position $(\theta,\phi)$ on a sphere of radius
$\sim 16$~Mpc centred on the origin for protons of initial energy
(a) $E_0=10^{19}$~eV and (b) $E_0=10^{20}$~eV injected
isotropically at the origin in the wall/void model of the IGMF
discussed in the text.  Gray-scale is logarithmic in $g(E)$ (see
text for details).}
\label{Fig6}
\end{figure*}

Results based on the average time per unit volume in $10^6$
simulations are shown for two initial energies $10^{19}$~eV and
$10^{20}$~eV in Fig.~\ref{Fig6}.  The gray-scale shows the
enhancement factor
\begin{equation}
g(\theta,\phi)=4\pi d^2c\tau(\theta,\phi)/V.
\label{eq:enhance}
\end{equation}
The strong peaks at $(0^\circ,0)$ and $(180^\circ,0)$ show that
despite the presence of the turbulent field component,
cross-field diffusion is not strong enough to give rise to
significant fluxes far away from the regular field threading the
source, at least for source distances as small as 16~Mpc for this
field topology, and we note that Stanev et
al.~\cite{StanevSeckelEngel2001} arrived at a similar conclusion
about cross-field diffusion.  The peaks in $g(\theta,\phi)$ shown
in Fig.~\ref{Fig6} are at a level $g(\theta,\phi) \sim 10^3$ over
a disk of radius $\sim 5^\circ$ ($\sim 1.4$~Mpc) outside of which
$g(\theta,\phi) \ll 1$, indicating that if the magnetic field
topology is such that the field lines connect the vicinity of M87
(within $\sim 1$~Mpc) to our Galaxy (within $\sim 1$~Mpc) then
cosmic rays from M87 should clearly be observed.  The average
ages are $\bar{t} \sim 200$~Mpc/c ($E_0=10^{19}$eV) and $\bar{t}
\sim 100$~Mpc/c ($E_0=10^{20}$eV) indicating their final energies
would be $\sim 8 \times 10^{18}$eV and $\sim 5 \times 10^{19}$eV
respectively.

\section{Conclusion}

We find the unresolved nuclear core of M87 to be consistent with
it being a mis-aligned BL Lac object.  In the context of the
Synchrotron Proton Blazar model, M87 could be either an HBL or an
LBL.  In both cases, we predict gamma-ray emission at levels
detectable by GLAST and MAGIC, and possibly by VERITAS.  Neutrino
detection is not expected, except possibly during an extreme
flare in the LBL case.

Ultra-high energy neutrons produced in pion photoproduction
interactions will escape from the host galaxy where they decay
into protons.  Because cosmic rays are deflected in the IGMF,
protons resulting from neutrons emitted in all directions can
contribute to the observed UHECR.  Hence, even though the
electromagnetic radiation we observe from the M87 jet is not
significantly Doppler boosted in energy, the cosmic ray output
will be.  By this mechanism, the unresolved nuclear core of the
M87 jet could emit UHECR with energies up to at least $\sim 3
\times 10^{20}$~eV.

We predict the UHECR output from M87 to be at a level such that
if UHECRs travelled in straight lines they would give an average
intensity at Earth a factor $\sim$20 below that observed.  For a
constant UHECR output and simple isotropic diffusion with a
scattering mean free path $\lambda$ less than the distance $d$ to
M87, the predicted intensity would be enhanced by a factor
$3d/\lambda$ relative to straight line propagation.  This
enhancement is accompanied by increased travel times, and so
higher energy losses in the CMBR.  Nevertheless, taking
reasonable scattering mean free paths we find that for simple
diffusion models protons from decay of ultra-high energy neutrons
produced by pion photoproduction in the M87 jet could easily
account for all the observed UHECR.  

In perhaps more realistic wall/void models of IGMF structure, M87
would only be a source of the observed UHECR if the topology of
the IGMF between M87 and our Galaxy is favourable.
Note, however, that because of its very high black hole mass M87
was probably much more active at earlier times than at present.
Many objects exhibit a high state for $\sim$5\% of the time, and 
since we estimate the average travel time of UHECR from M87 to
Earth to be a factor $\sim$5--10 times the light propagation time
it is indeed possible that the UHECR observed now were emitted
when M87 was in a high state and possibly a more powerful cosmic ray
source.  Similarly, if M87 was in the recent past a powerful FR-II radio
galaxy, it could have been possible to accelerate UHECR
at shocks in the giant radio lobes where the jets terminate
as in the model of Rachen and Biermann~\cite{RachenBiermann1993},
and these could also contribute to the observed UHECR.

\section*{Acknowledgments}

We thank Todor Stanev for very helpful comments. This work was
supported by a Discovery Project grant from the Australian Research Council
and a grant from the University of Adelaide to RJP.
AR thanks the Bundesministerium f\"ur Bildung und Forschung for financial
support through DESY grant Verbundforschung 05CH1PCA/6.

\newpage

\end{document}